\documentclass[
  ,final            
  ]
  {aipproc}
  
  \usepackage{epsfig}

\layoutstyle{6x9}

\def\Z{{\mathchoice {\hbox{$\sf\textstyle Z\kern-0.4em Z$}}
{\hbox{$\sf\textstyle Z\kern-0.4em Z$}}
{\hbox{$\sf\scriptstyle Z\kern-0.3em Z$}}
{\hbox{$\sf\scriptscriptstyle Z\kern-0.2em Z$}}}}
\def\abs#1{\left| #1\right|}

\def\square{\kern1pt\vbox{\hrule height 1.2pt\hbox{\vrule width 1.2pt
   \hskip 3pt\vbox{\vskip 6pt}\hskip 3pt\vrule width 0.6pt}
   \hrule height 0.6pt}\kern1pt}

\def\mn{{\mu\nu}}

\def\e{\,{\rm e}}

\newcommand{\be}{\begin{equation}}
\newcommand{\ee}{\end{equation}\noindent}
\newcommand{\bear}{\begin{eqnarray}}
\newcommand{\ear}{\end{eqnarray}\noindent}
\newcommand{\benn}{\begin{enumerate}}
\newcommand{\enn}{\end{enumerate}}

\newcommand{\no}{\noindent}
\def\non{\nonumber}

\begin{document}

\title{Pair creation in inhomogeneous fields from worldline instantons
\footnote{Talk given by C. S. at \emph{X Mexican Workshop on Particles and Fields},
Morelia, Mexico, Nov. 6 -- 12, 2005 (to appear in the conference proceedings).
}}


\author{Gerald V. Dunne}{
  address={Department of Physics, University of Connecticut, Storrs, Connecticut 06269-3046, USA}
}

\author{\underline{Christian Schubert}}{
  address={Instituto de F\1sica y Matem\'aticas, Universidad Michoacana de San Nicol\'as
  de Hidalgo, Ed. C-3, C.U., C.P. 58040, Morelia, Michoac\'an, M\'exico}
 } 
 

\begin{abstract}
We show how to do semiclassical nonperturbative computations within the
 worldline approach to quantum field theory using ``worldline instantons''.
 These worldline instantons are classical solutions to the Euclidean
 worldline loop equations of motion, and are closed spacetime loops
 parametrized by the proper-time. Specifically, we compute the imaginary
 part of the one loop effective action in scalar and spinor QED using
 worldline instantons, for a wide class of inhomogeneous electric field backgrounds. 
 \end{abstract}

\maketitle


\vspace{15pt}

\section{Introduction}

\vspace{10pt}

\no
As is well-known, quantum field theory allows the spontaneous
creation of electron-positron pairs from vacuum in external electric fields. 
This effect has been considered already in the early days of quantum electrodynamics
\cite{sauter, eulhei}, and Schwinger  
used effective action methods 
to obtain a simple closed-form
expression for the production rate in the constant field case \cite{schwinger51}.
Although spontaneous pair creation
is of potential interest for many branches of physics 
the chances for its direct experimental verification hitherto seemed 
very remote. This is  due to the exponential smallness of the production rate 
for field strengths below the critical value
$E_c = {m^2c^3\over e\hbar} (1.3 \times 10^{18} V/m)$, which is far above 
the electric fields which can be produced in the laboratory macroscopically. 
However, given the rapid progress of laser technology it seems not any more
impossible that in the near future pair production might be observable in laser fields.
Both the optical laser system POLARIS \cite{heinEA} under construction at the Jena
high-intensity laser facility and the X-ray free electron lasers to be constructed at SLAC
\cite{LCLS} and DESY \cite{XFEL} are expected to reach laser field strengths missing
$E_c$ only by a few orders of magnitude. Moreover, it has been argued that for focused
laser pulses substantial pair creation should set in already somewhat below critical
field strength \cite{bnmp}. 

Laser fields cannot usually be treated in the constant field approximation,
and this is particularly true for the pair creation process due to its nonperturbative
nature. Much effort has gone into developing methods for the
calculation of pair creation rates in inhomogeneous fields, mostly based on WKB
\cite{keldysh,nikrit,breitz,narnik,popmar,kimpag}. 
In this talk I will present a substantially different approach \cite{wlinst1} based
on Feynman's worldline path integral formalism \cite{feynman50,feynman51}
and  work done by Affleck et al. in 1982 for the
constant field case in scalar QED
\cite{afalma}.
The path integral representing the imaginary part of the
one loop effective action is calculated in a semiclassical approximation around a closed classical 
``instanton'' trajectory. 
The worldline action evaluated on
this solution directly gives the Schwinger exponent of the imaginary part of the effective
action in the weak field approximation.      

\vspace{10pt}

\section{Pair creation in electric fields}

Let us start with
Schwinger's well-known formula for the imaginary parts of the scalar and spinor QED
effective Lagrangians in a constant electric field
(at one loop) \cite{schwinger51}:

\bear
{\rm Im} {\cal L}^{(1)}_{\rm scalar}(E) &=&  \frac{m^4}{16\pi^3}
\beta^2\, \sum_{n=1}^\infty \frac{(-1)^{n-1}}{n^2}
\,\exp\left[-\frac{\pi n}{\beta}\right]
\non\\
{\rm Im} {\cal L}^{(1)}_{\rm spinor}(E) &=&  \frac{m^4}{8\pi^3}
\beta^2\, \sum_{n=1}^\infty \frac{1}{n^2}
\,\exp\left[-\frac{\pi n}{\beta}\right]\non\\
\label{ImL1}
\ear
where $\beta = {eE\over m^2}$. The first term in each series gives (up to a factor of $2$)
directly the total pair-production rates per volume per time \cite{nikishov}. The higher order 
($n\geq 2$) terms are statistics dependent and contain the information on the 
coherent production of $n$ pairs by the field. All terms are exponentially suppressed for
$E <  E_{\rm cr}$. 
Higher loop corrections have also been considered \cite{lebrit,rss}. They can be neglected for
subcritical fields as far as the total pair production rate is concerned \cite{leipzig}.

The formulas (\ref{ImL1}) are commonly derived from the standard propertime
representation for the one-loop effective Lagrangian. For spinor QED, this is
the Euler-Heisenberg Lagrangian \cite{eulhei}:

\bear
{\cal L}_{\rm spinor}^{(1)}(E) = - {1\over 8\pi^2}
\int_0^{\infty}{dT\over T^3}
\,\e^{-m^2T} 
\Biggl[
{eET\over \tan(eET)} + {1\over 3}(eET)^2 -1
\Biggr]
\label{L1spin}
\ear
The $n$th term in the series for ${\rm Im}{\cal L}_{\rm spinor}$ in (\ref{ImL1})
is generated y the pole at $T = {n\pi\over eE}$ of the integrand in (\ref{L1spin}).

\section{Schwinger's formula from worldline instantons}

Feynman in 1950 \cite{feynman51}
presented, ``as an alternative to the formulation of second quantization'',
a formula representing the scalar QED effective action $\Gamma_{\rm scalar}[A]$ in terms of
first-quantized worldline path integrals. In the quenched approximation
(i.e. with only one scalar loop but any number of photons) it reads

\bear
\Gamma_{\rm scalar}^{(\rm quenched)}[A] =
\int d^4x\, {\cal L}^{({\rm quenched})}_{\rm scalar}[A]
=
\int_0^{\infty}{dT\over T}\,{\rm e}^{-m^2T}
{\displaystyle \int_{x(T)=x(0)}}{\cal D}x(\tau)
\, e^{-S[x(\tau)]}
\label{feynman}
\ear
Here the path integral is over all closed loops in spacetime with a given period $T$
in the proper-time of the loop scalar. The worldline action 
$S=S_0+S_e+S_i$ has three parts,

\bear
S_0 &=& \int_0^T d\tau\, {\dot x^2\over 4}  \nonumber\\
S_e &=& ie\int_0^T \dot x^{\mu}A_{\mu}(x(\tau))
\nonumber\\
S_i &=&
-{e^2\over 8\pi^2}\int_0^Td\tau_1\int_0^Td\tau_2 {\dot x(\tau_1)\cdot\dot x(\tau_2)\over
(x(\tau_1)-x(\tau_2))^2}\nonumber\\
\label{Pwlintegrand}
\ear
Of these $S_e$ incorporates the interaction with the external field, while
$S_i$ takes all internal photon exchanges in the loop
into account. 

Feynman generalized this ``worldline representation'' to spinor QED
in \cite{feynman51}. Although it
has always been considered an interesting alternative to standard
quantum field theory, only in recent years it 
has gained some popularity as a calculational tool. 
Presently there exist, among others, the following approaches to
the calculation of this type of path integral:

\begin{itemize}

\item
The ``string-inspired'' approach 
\cite{polyakov,berkos,strassler,rss,fhss} (see \cite{report} for a review) 
which aims at an analytical calculation using appropriate wordline
Green's functions.

\item
Semiclassical calculation using a stationary phase approximation 
\cite{afalma}.

\item
Variational methods \cite{ars}.

\item
Numerical calculation using 
Monte Carlo methods \cite{gies,schsta}. This approach in principle
applies to effective actions in arbitrary backgrounds, and has
been applied to the pair creation process in \cite{giekli}.

\end{itemize}

We will expand here on the first approach, which is inspired by instanton
methods in field theory. The idea is to calculate   
${\rm Im}{\cal L}_{\rm scalar}$ for weak fields using an extremal trajectory of the
worldline path integral for a stationary phase approximation.
In \cite{afalma} it was shown that for the case of a constant electric field in the $z$ direction this 
extremal action trajectory (``worldline instanton'') is given by a circle in the
(euclidean) $t-z$ plane:
  
\bear
x_{\rm extremal}(\tau) = {m\over eE}\bigr(0,0,{\rm cos}(2\pi \tau),{\rm sin}(2\pi \tau)\bigl)
\qquad\qquad \bigl(T=1\bigr)
\label{wlinst}
\ear
\no
At leading order in the stationary phase approximation, the exponent of the imaginary part of 
the effective Lagrangian is given by the worldline action of this
trajectory,

\bear
{\rm Im}{\cal L}_{\rm scalar}^{({\rm quenched})}(E)  \sim {\rm e}^{-S[x_{\rm extremal}]}\label{Sextremal}
\ear
This is easily evaluated to be

\bear
(S_0+S_e)[x_{\rm extremal}] = \pi {m^2\over eE}, \qquad
S_i[x_{\rm extremal}] = - \alpha\pi \, . \label{Sinst}
\ear
The contribution of $S_0+S_e$ just reproduces the first of the exponentials in
Schwinger's one-loop formula (\ref{ImL1}), while the higher order ones are generated by the
``multi-instantons'' where the same circle is traversed $n$ times. 
More surprising is the simplicity of the $S_i$ term, which 
represents the contribution of all the higher loop corrections
involving arbitrary photon exchanges in the loop. According to
\cite{afalma} this is the exact all-orders result in the weak field limit, 
including renormalization effects:

\bear
{\rm Im}\,{\cal L}_{\rm scalar}^{({\rm quenched})}(E)\,\, \,\,
\stackrel{\beta\to 0}{\simeq}\,\,\,\, {m^4\over 8\pi^3}\beta^2\,
{\rm exp}\biggl(-{\pi\over\beta}+\alpha\pi \biggr)
\label{master}
\ear
At the two-loop level this remarkable formula has
been independently
confirmed \cite{ds2_colima}. 

\section{Inhomogeneous background fields}

Despite of the simplicity and elegance of this worldline instanton approach
it appears that the work of \cite{afalma} has never been extended either to
spinor QED or to more general backgrounds. As we will now show, at least 
at the one-loop level the method generalizes to a large class of inhomogeneous
backgrounds straightforwardly. Let us return to Feynman's formula (\ref{feynman}),
omitting the  photon exchange term $S_i$:  
                                                                                                                                   
\bear
\Gamma_{\rm scalar}^{({\rm 1-loop})} [A] &=&
\int_0^{\infty}{dT\over T}\, \e^{-{m^2}T}
\int {\cal D}x 
\, \e^{-\int_0^Td\tau 
\bigl({\dot x^2\over 4} +ieA\cdot \dot x \bigr)}
\label{Gammaoneloop}
\ear
Rescaling $\tau = Tu$, this becomes

\bear
\Gamma_{\rm scalar}^{({\rm 1-loop})}[A] &=&
\int_0^{\infty}{dT\over T}\, \e^{-{m^2}T}
\int {\cal D}x 
\, \e^{-\Bigl({1\over T}\int_0^1du \,
\dot x^2 +ie\int_0^1du A\cdot \dot x 
\Bigr) }
\label{Gammarescaled}
\ear
The $T$ integral has a
stationary point at 

\bear
T_0^2 &=& {\int du\,\dot x^2\over m^2}
\label{T0}
\ear
leading to

\bear
{\rm Im}\, \Gamma^{({\rm 1-loop})}_{\rm scalar} &=&
{1\over m}\sqrt{2\pi\over T_0}
\,{\rm Im} \int {\cal D}x \, 
\e^{-\Bigl(m\sqrt{\int du\, \dot x^2} 
+ie\int_0^1 du A\cdot \dot x
\Bigr)}
\label{ImGamma}
\ear
The new worldline action,

\bear
S &=& m\sqrt{\int du \,\dot x^2} + ie \int_0^1du \, A\cdot \dot x
\label{newaction}
\ear
is stationary if

\bear
m{\ddot x_{\mu}\over \sqrt{\int du\,\dot x^2}} &=& ie F_{\mn}\dot x_{\nu}
\label{eom}
\ear
Contracting with $\dot x^{\mu}$ yields 
$\dot{x}^2={\rm constant}\equiv a^2$
so that

\bear
m\ddot x_{\mu} &=& iea F_{\mn}\dot x_{\nu}
\label{eoma}
\ear

\no
Now specialize to a {\sl time-dependent electric field directed in the $x_3$ direction}.  Choose a gauge
where

\bear
A_3=A_3(x_4) \!; \quad A_\mu=0\,\,\,{\rm for}\,\, \mu\neq 3.
\label{A3temporal}
\ear
\no
Since $F_{\mu 1}=F_{\mu 2}=0$, the stationarity conditions together with
the periodicity imply that $x_1$ and $x_2$ must be constant. 
Hence one is down to an equation for $x_3$ and $x_4$:
\bear
\ddot{x}_{3} &=& \frac{i e a}{m}\, F_{34}\, \dot{x}_{4},
\qquad
\ddot{x}_{4}= -\frac{i e a}{m}\, F_{34}\, \dot{x}_{3}\, .
\non\\
\label{equF}
\ear
In terms of $A_3$ this can be further reduced to

\bear
\dot{x}_3&=&-\frac{i e a}{m}\, A_3(x_4), \qquad
\abs{\dot{x}_4}= a\, \sqrt{1+\left(\frac{e\, A_3(x_4)}{m}\right)^2}.
\non\\
\label{equA}
\ear
As an example, let us consider the following
single-pulse electric background
\cite{popmar,narnik},

 \bear
 E(t)=E\, {\rm sech}^2(\omega\,t)
 \label{singlepulse}
 \ear
For this background the solution of (\ref{equF}) turns out to be very simple:

\bear
x_3(u)&=&-\frac{1}{\omega}\,\frac{1}{\sqrt{1+\gamma^2}} \, {\rm arcsinh}\left[\gamma\, \cos\left(2 n\pi u\right)\right]\non\\
x_4(u)&=&\frac{1}{\omega}\, \arcsin\left[\frac{\gamma}{\sqrt{1+\gamma^2}}\, \sin\left(2 n \pi \, u\right)\right]
\non\\
\label{solsinglepulse}
\ear
Here 
$\gamma\equiv \frac{m\omega}{eE}$
denotes the ``adiabaticity parameter'' \cite{keldysh} and the integer
$n\in {\bf Z}^+$ counts the number of times the closed path is traversed.
The worldline action (\ref{newaction}) evaluated on this instanton is

\bear
S_0&=&n\, \frac{m^2 \pi}{e E}\left(\frac{2}{1+\sqrt{1+\gamma^2}}\right)
\label{actionsinglepulse}
\ear
In fig. 1 we plot the the instanton trajectories for various values of the parameter $\gamma$. 
In the static limit $\gamma\to 0$ we recover the circular paths (\ref{wlinst}) of the constant field case.
In the short-pulse limit $\gamma\to\infty$ the instantons shrink in size and become elongated. 
Thus the instanton action decreases with increasing $\gamma$, leading to a
local enhancement of
the pair creation rate as compared to the case of a constant field with magnitude $E$.    

\begin{figure}[h]
\centerline{\includegraphics[height=.4\textheight]{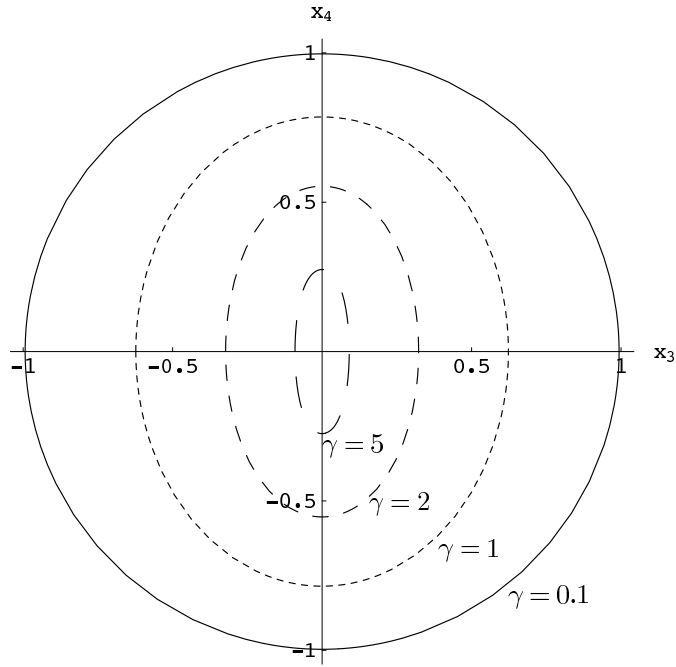}}
  \caption{Plot of the worldline instanton paths (\ref{solsinglepulse}) in the $(x_3,x_4)$ plane
  for the case of
  a time-dependent electric field $E(t) = E{\rm sech}^2(\omega t)$. The paths are shown for
  various values of the adiabaticity parameter $\gamma$. $x_{3,4}$ have been expressed
  in units of ${m\over eE}$.}
\end{figure}

The case of a {\sl spatially} inhomogeneous electric field in the $z$ direction can be
treated completely analogously. The (euclidean) gauge potential can be chosen as

\bear
A_4 = A_4(x_3); \qquad A_{\mu} = 0 \quad {\rm for}\,\, \mu \ne 4 \, .
\label{Aspatially}
\ear
\vspace{-5pt}

\no
This leads to instanton equations differing from eqs. (\ref{equA})  simply by the interchange
$3\leftrightarrow 4$. 

As an example, let us consider the spatial analogue of 
(\ref{singlepulse}), i.e. a single-bump electric field depending only on $x_3$:

\bear
 E(x_3)=E{\rm sech}^2(kx_3)
\label{singlebump}
\ear
The instanton solutions are obtained from the single-pulse ones (\ref{solsinglepulse})
{\it mutatis mutandis}:

\bear
x_3(u) &=& {m\over eE} {1\over\tilde \gamma}{\rm arcsinh}
\biggl( {\tilde\gamma\over\sqrt{1-\tilde\gamma^2}}
\sin(2\pi nu) 
\biggr)\non\\
x_4(u) &=& {m\over eE} {1\over\tilde\gamma\sqrt{1-\tilde\gamma^2}}
{\rm arcsin}\,(\tilde\gamma \cos(2\pi nu)) \non\\
\label{solsinglebump}
\ear
with $\tilde\gamma = {mk\over eE}$. The stationary action is

\bear
S_0 &=& n{m^2\pi\over eE}
\biggl ( {2\over 1+\sqrt{1-\tilde\gamma^2}}\biggr)
\label{action singlebump}
\ear

\vspace{10pt}

These solutions are plotted in fig. 2 for various
values of $\tilde\gamma$. Note that again they reduce to the constant field circles
for $\tilde\gamma \to 0$, but they grow in size with increasing $\tilde\gamma$ and
become infinitely large when $\tilde\gamma\to 1$. Physically, this is where the width of
the electric field becomes too small for a virtual pair to extract from it the energy 
necessary to turn real. 

\begin{figure}[h]
\centerline{\includegraphics[height=.4\textheight]{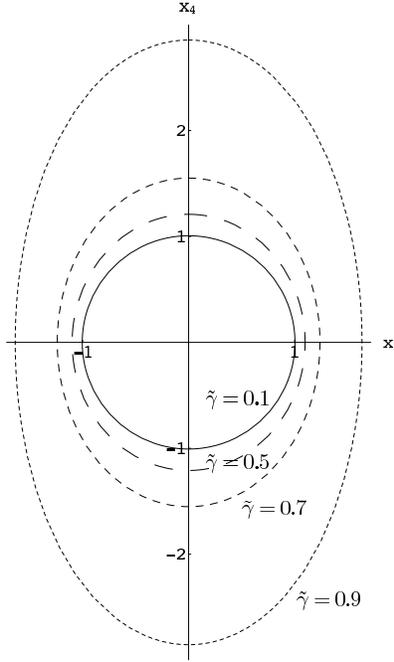}}
  \caption{Plot of the instanton paths (\ref{solsinglebump})  for the case of
  a space-dependent field $E(x)=E{\rm sech}^2(kx)$ for
  various values of the adiabaticity parameter $\tilde\gamma$.}
\end{figure}

\no
The instanton action increases with increasing $\tilde\gamma$, leading
to a lower local pair production rate as compared to the constant field case.
Thus we see a rule emerging here which we believe holds quite generally:

\bigskip

\begin{itemize}

\item
{\it Temporal inhomogeneity increases the pair production rate}

\medskip

\item
{\it Spatial inhomogeneity decreases the pair production rate}

\end{itemize}

\section{The spinor loop case}

A path integral representation for the one loop effective action in spinor QED
\cite{feynman51}
can be obtained from the scalar QED one                                                                                           (\ref{Gammaoneloop}) by multiplication by a global factor of $-{1\over 2}$ and
the insertion of the following ``spin factor'' $S[x,A]$,
 
\vspace{-5pt} 
 
\bear
S[x,A] &=& {\rm tr}_{\Gamma} {\cal P}
\e^{{i\over 2}e\sigma^{\mu\nu}
\int_0^Td\tau F_{\mu\nu}(x(\tau))}
\label{spinfactor}
\ear
For the two-dimensional background fields considered above the path ordering has no
effect, since the $F_{\mn}(x(\tau))$'s at different proper-times commute. The spin factor
then reduces to

\vspace{-15pt}

\bear
S[x,A] = 4\cos \Bigl[
eT \int_0^1 du\, E(x(u)) \Bigr]
\label{spinspecial}
\ear
Since the exponent of the spin factor (\ref{spinfactor}) is purely imaginary it affects neither
the determination of the stationary point $T_0$ nor the instanton equations (\ref{eom}). 
Therefore it remains only to evaluate the spin factor on the instanton solutions 
for the scalar loop. For the class of backgrounds of the form $A_3(x_4)$ or $A_4(x_3)$
considered above the result turns out to be \cite{wlinst1}

\vspace{-12pt}

\bear
S[x,A] &=& 4(-1)^n
\label{spinresult}
\ear
Thus at least for this class of inhomogeneous backgrounds we find 
that 
${\rm Im} {\cal L}^{(1)}_{\rm scalar}$ and 
${\rm Im} {\cal L}^{(1)}_{\rm spinor}$
differ by the the same simple
sign changes as in the constant field case, eq.(\ref{ImL1}).

\section{Summary, work in progress}

We have shown that the worldline instanton approach holds considerable
promise as a tool for calculating pair creation rates in inhomogeneous backgrounds
in scalar and spinor QED. Although the class of backgrounds which we have considered
here is also amenable to a treatment by WKB methods \cite{breitz,narnik,popmar,kimpag}
the worldline approach
offers a number of distinct advantages: (i) it bypasses the momentum space
integrals which usually would have to be done in this type
of calculation
(ii) it uses proper-time instead of time, which seems
more natural in the treatment of an intrinsically relativistic effect such as pair creation 
(iii) it provides a framework for the inclusion of radiative corrections.

Clearly we have not fully explored here the potential of this method.
The instanton equations in the form (\ref{eom}) generalize immediately
to the case of an arbitrary electromagnetic background field. While
closed-form solutions can be expected only in special cases,
the numerical integration of these equations poses no problems in
principle. However, the instanton provides only the exponents of the Schwinger exponentials.  
We have not discussed here the
prefactors of the Schwinger exponentials (\ref{ImL1}), which in the present
approach involve the determinant of fluctuations around the worldline
instanton. In the constant field case the determinant computation is
straightforward, since the fluctuation problem is Gaussian \cite{afalma}.
As will be shown in a forthcoming paper \cite{wlinst2} for
inhomogeneous background fields the determinant can be computed using the
Gelfand-Yaglom technique, since the fluctuation operator is an ordinary
differential operator, depending only on the proper-time.


\begin{theacknowledgments}
We are very grateful to Don Page for helpful correspondence. We acknowledge the
support of the NSF US-Mexico Collaborative Research Grant 0122615.
\end{theacknowledgments}

\vfill\eject





\begin{thebibliography}{99}

\bibitem{sauter}
F. Sauter, \emph{Z. Phys.} \textbf{69}, 742 (1931).

\bibitem{eulhei}
W. Heisenberg and H. Euler, \emph{Z. Phys.} \textbf{98}, 714 (1936).

\bibitem{schwinger51}
J. Schwinger, \emph{Phys. Rev.} \textbf{82}, 664 (1951).

\bibitem{heinEA}
J. Hein et. al., \emph{Appl. Phys. B} \textbf{79}, 419 (2004).

\bibitem{LCLS}
Linac Coherent Light Source, 
http://www-ssrl.slac.stanford.edu/lcls/

\bibitem{XFEL}
European X-Ray Laser Project XFEL,
http://xfel.desy.de/

\bibitem{bnmp}
S.S. Bulanov, N.B. Narozhny, V.D. Mur, and V.S. Popov,
\emph{Phys. Lett. A} \textbf{330}, 1 (2004) [arXiv:hep-ph/0403163].

\bibitem{keldysh}
L.V. Keldysh, \emph{Sov. Phys. JETP} \textbf{20}, 1307 (1965).

\bibitem{nikrit}
A.I. Nikishov and V.I. Ritus, \emph{Sov. Phys. JETP} \textbf{25}, 1135 (1967).

\bibitem{breitz}
E. Brezin and C. Itzykson, \emph{Phys. Rev. D}  \textbf{2}, 1191 (1970).

\bibitem{narnik}
N.B. Narozhnyi and A.I. Nikishov, \emph{Yad. Fiz.} \textbf{11}, 1072 (1970)
[\emph{Sov. J. Nucl. Phys.} \textbf{11}, 596 (1970)].

\bibitem{popmar}
V. S. Popov,
\emph{Sov. Phys. JETP} \textbf{34}, 709 (1972);
\emph{Sov. Phys. JETP} \textbf{35}, 659 (1972);
V.~S.~Popov and M.~S.~Marinov,
\emph{Yad.\ Fiz.}  \textbf{16}, 809 (1972) [\emph{Sov. J. Nucl. Phys.} \textbf{16}, 449 (1973)].

\bibitem{kimpag}
S.P. Kim and D.N. Page, 
\emph{Phys. Rev. D}  \textbf{65}, 105002 (2002) [arXiv:hep-th/0005078];
\emph{Phys. Rev. D} \textbf{63}, 065020 (2006) [arXiv:hep-th/0301132].

\bibitem{wlinst1}
G.V. Dunne and C. Schubert,
\emph{Phys. Rev. D} \textbf{72}, 105004 (2005) [arXiv:hep-th/0507174].

\bibitem{feynman50}
R.P. Feynman, \emph{Phys. Rev.}  \textbf{80}, 440 (1950).

\bibitem{feynman51}
R.P. Feynman, \emph{Phys. Rev.}  \textbf{84}, 108 (1951).

\bibitem{afalma}
I.K. Affleck, O. Alvarez, and N.S. Manton, \emph{Nucl. Phys. B}   \textbf{197}, 509 (1982).

\bibitem{nikishov}
A.I. Nikishov, \emph{Sov. Phys. JETP}  \textbf{30}, 660 (1970).

\bibitem{lebrit}
S.L. Lebedev and V.I. Ritus, \emph{Sov. Phys. JETP}  \textbf{59}, 237 (1984)
[\emph{Zh. Eksp. Teor. Fiz.} \textbf{86}, 408 (1984)].

\bibitem{rss}
M. Reuter, M.G. Schmidt, and C. Schubert, \emph{Ann. Phys. (N.Y.)} \textbf{259}, 313 (1997)
[arXiv:hep-th/9610191].

\bibitem{leipzig}
G.V. Dunne and C. Schubert, 
\emph{Fifth workshop on quantum field theory under the influence of 
external conditions, Leipzig 2001}, edited by K. Bordag,
\emph{Int. J. Mod. Phys. A} \textbf{17}, 956 (2002).

\bibitem{polyakov}
A. M. Polyakov, \emph{Gauge Fields and Strings},  Harwood Publ., Chur, 1987.

\bibitem{berkos}
Z. Bern and D.A. Kosower, \emph{Nucl.\ Phys.\ B} \textbf{379}, 451 (1992).

\bibitem{strassler}
M. J. Strassler, \emph{Nucl.\ Phys.\ B} \textbf{385}, 145 (1992)
[arXiv:hep-ph/9205205].

\bibitem{fhss}
D. Fliegner, P. Haberl, M.G. Schmidt, and C. Schubert, 
\emph{Ann. Phys. (N.Y.)}  \textbf{264}, 51 (1998)
[arXiv:hep-th/9707189].

\bibitem{report}
C. Schubert, \emph{Phys.\ Rept.} \textbf{355}, 73 (2001)
[arXiv:hep-th/0101036].

\bibitem{ars}
C. Alexandrou, R. Rosenfelder, and W. Schreiber, 
\emph{Phys. Rev. D} \textbf{62} 085009 (2000) 
[arXiv:hep-th/0003253].

\bibitem{gies}
H. Gies and K. Langfeld, \emph{Nucl. Phys. B} \textbf{613}, 353 (2001)
[arXiv:hep-ph/0102185].

\bibitem{schsta}
M.G. Schmidt and I.O Stamatescu, \emph{Mod. Phys. Lett.} \textbf{18}, 1499 (2003).

\bibitem{giekli}
H. Gies and K. Klingm\"uller,
\emph{Phys. Rev. D}  \textbf{72}, 065001 (2005) [arXiv:hep-ph/0505099].

\bibitem{ds2_colima}
G.V. Dunne and C. Schubert, \emph{IX Mexican Workshop on Particles and Fields, Colima 2003}
[arXiv:hep-th/0409021].

\bibitem{wlinst2}
G.V. Dunne, H. Gies, C. Schubert, and Q.-H. Wang, 
\emph{Phys. Rev. D}  \textbf{73}, 065028 (2006) [arXiv:hep-th/0602176]
(completed after the workshop).

\end{thebibliography}




\end{document}